\documentclass[aps,prb,twocolumn,amsfonts,amsmath,amssymb,superscriptaddress]{revtex4-1}
\usepackage{graphicx}
\usepackage{dcolumn}
\begin{document}

\title{Identification of excitons, trions and biexcitons in single-layer WS$_2$}

\author{Gerd Plechinger}
\email{gerd.plechinger@physik.uni-r.de
  , Phone:
  +49-941-9432053, Fax: +49-941-9434226}
\affiliation{Institut f\"ur Experimentelle und Angewandte Physik,
Universit\"at Regensburg, D-93040 Regensburg, Germany}
\author{Philipp Nagler}
\affiliation{Institut f\"ur Experimentelle und Angewandte Physik,
Universit\"at Regensburg, D-93040 Regensburg, Germany}
\author{Julia Kraus}
\affiliation{Institut f\"ur Experimentelle und Angewandte Physik,
Universit\"at Regensburg, D-93040 Regensburg, Germany}
\author{Nicola Paradiso}
\affiliation{Institut f\"ur Experimentelle und Angewandte Physik,
Universit\"at Regensburg, D-93040 Regensburg, Germany}
\author{Christoph Strunk}
\affiliation{Institut f\"ur Experimentelle und Angewandte Physik,
Universit\"at Regensburg, D-93040 Regensburg, Germany}
\author{Christian Sch\"uller}
\affiliation{Institut f\"ur Experimentelle und Angewandte Physik,
Universit\"at Regensburg, D-93040 Regensburg, Germany}
\author{Tobias Korn}
\affiliation{Institut f\"ur Experimentelle und Angewandte Physik,
Universit\"at Regensburg, D-93040 Regensburg, Germany}

\begin{abstract}
Single-layer WS$_2$ is a  direct-gap semiconductor showing strong excitonic photoluminescence features in the visible spectral range. Here, we present temperature-dependent photoluminescence measurements on mechanically exfoliated single-layer WS$_2$, revealing the existence of neutral and charged excitons at low temperatures as well as at room temperature. By applying a gate voltage, we can electrically control the ratio of excitons and trions and assert a residual n-type doping of our samples. At high excitation densities and low temperatures, an additional peak at energies below the trion dominates the photoluminescence, which we identify as biexciton emission.
\end{abstract}
\maketitle   
In recent years, semiconducting, atomically thin transition metal dichalcogenides (TMDCs) like MoS$_2$, MoSe$_2$, WSe$_2$ and WS$_2$, have emerged as highly interesting materials for the scientific community due to their extraordinary optical~\cite{HeinzPRL10} and electrical properties~\cite{Kis_NatNano10}, including coupled spin-valley effects~\cite{Mak12} and photovoltaic applications~\cite{Mueller14}. These molecular layers show strong photoluminescence (PL) peaks in the visible and near-infrared spectral range, as they experience a transition from an indirect gap in bulk and few-layer samples to a direct gap in the single-layer regime~\cite{Splen_Nano10}. The spatial confinement of carriers in a two-dimensional layer and the weak dielectric screening lead to unusually strong excitonic effects~\cite{Chei12,Malic14},   even at room temperature. High exciton binding energies of the order of 0.5\,eV have been reported for single-layer WS$_2$ \cite{Chernikov_Rydberg_PRL14,Zhang14,Cui15,Jonker15}. Besides the charge-neutral exciton (X), i.e., a bound state   of an electron and a hole, also charged excitons can be excited in the presence of residual excess charge carriers. These quasiparticles, called trions, consist either of two electrons and one hole (X$^-$) or one electron and two holes (X$^+$). By applying a gate voltage, one can  tune the spectral weight of charge-neutral excitons and trions in single-layer MoS$_2$~\cite{Mak13}, MoSe$_2$~\cite{Ross13}, WS$_2$~\cite{Cui15}  and WSe$_2$~\cite{Jones13}. Additional, lower-energy PL emission peaks  are observed in most single-layer TMDCs at low temperatures. These  have been attributed to surface-adsorbate-bound excitons in MoS$_2$~\cite{Plechinger12} and to crystal-defect-bound exciton states in single-layer diselenides~\cite{Wang14,Marie14}.
Given the large binding energy of the excitons, the formation of molecular states consisting of two excitons, so-called biexcitons~\cite{Kling07}, is to be expected in dichalcogenide single-layers. Biexciton PL emission should be at energies below the exciton emission due to the additional binding energy, in a similar energy range as defect-bound exciton emission. The signature of biexciton emission was recently observed in  PL measurements on WSe$_2$~\cite{Heinz15}. Thus, the origin of the lower-energy PL emission peaks in the other semiconducting TMDCs warrants close investigation.

In this work we report on low-temperature PL of mechanically exfoliated single-layer WS$_2$.  To date, only a few works exist which report on the observation of excitons and trions in the low-temperature PL spectrum of mechanically exfoliated~\cite{Hawrylak15,Mitioglu} single-layer WS$_2$. To the best of our knowledge, a thorough analysis of the temperature-dependent PL spectrum is still absent. In contrast to other semiconducting TMDCs, there is no consensus about the  assignment of the X and X$^-$ PL features in low-temperature PL of single-layer WS$_2$. The aim of this paper is to clarify those issues, and to provide insight into the nature of an additional low-energy peak in the PL spectrum, which is observable at low temperatures.  We identify the exciton and the trion peaks in the temperature range from 295\,K to 4\,K. Our interpretation of the PL spectra is substantiated by gate-dependent PL measurements which allow us to directly control the exciton-trion ratio. Finally, we utilize power-dependent and helicity-resolved PL measurements to show that the low-energy PL peak we observe stems from a superposition of defect-bound exciton and biexciton emission.

Our samples are mechanically exfoliated from bulk WS$_2$ crystals (2d semiconductors inc.) onto a polydimethylsiloxane (PDMS) stamp. Using an optical microscope, we can identify single-layer flakes of WS$_2$ on the PDMS stamp. We then transfer these flakes onto a p-doped Si chip with a 270\,nm SiO$_2$ capping layer, applying an all-dry deterministic transfer procedure~\cite{acg14}. For gate-dependent measurements, we stamp the flakes onto p-doped Si chips with 500\,nm thermal oxide and predefined metal contacts manufactured with e-beam lithography. We use the p-doped Si as a backgate. For low-temperature measurements, the samples are mounted in a He-flow cryostat. The cw lasers used for excitation are focussed with a 100x microscope objective onto the sample,  the emitted PL is collected by the same microscope objective (backscattering geometry) and guided into a spectrometer with a Peltier-cooled CCD chip. Unless otherwise noted, a 532 nm laser source is utilized. Helicity-resolved measurements are performed using a 561 nm laser, which allows for near-resonant excitation. Further experimental details are published in Ref.~\cite{Plechinger_SST14}.
\begin{figure}[t]%
\includegraphics*[width=\linewidth]{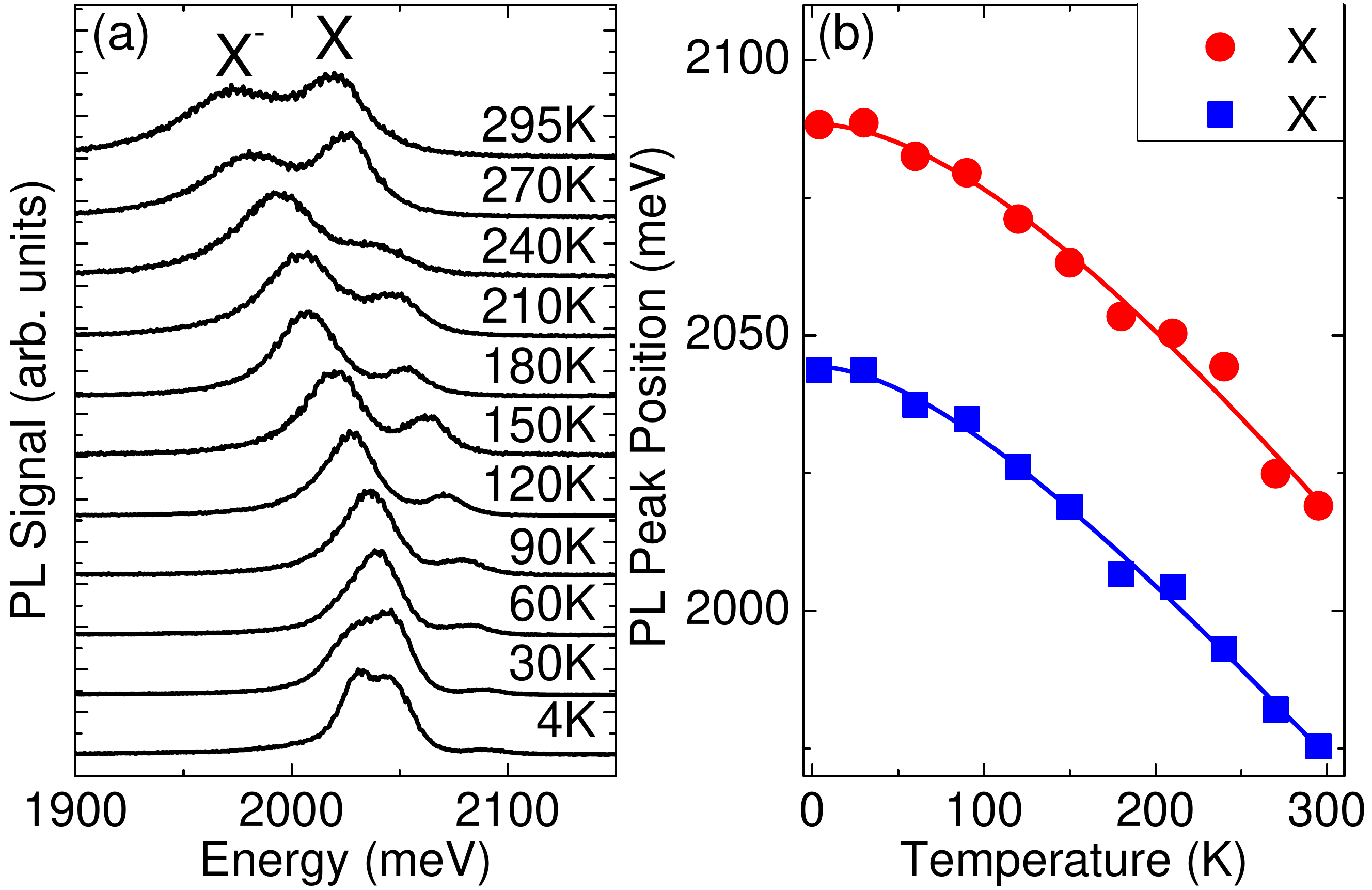}
\caption{%
  (a) Normalized PL spectra of single-layer WS$_2$ for different temperatures. (b) Exciton (X) and trion (X$^-$) PL peak energies as a function of temperature. The solid lines represent the fits to the experimental data following the Varshni equation.}
\label{fig:TempSeries}
\end{figure}
\begin{figure}[htb]%
\includegraphics*[width= \linewidth]{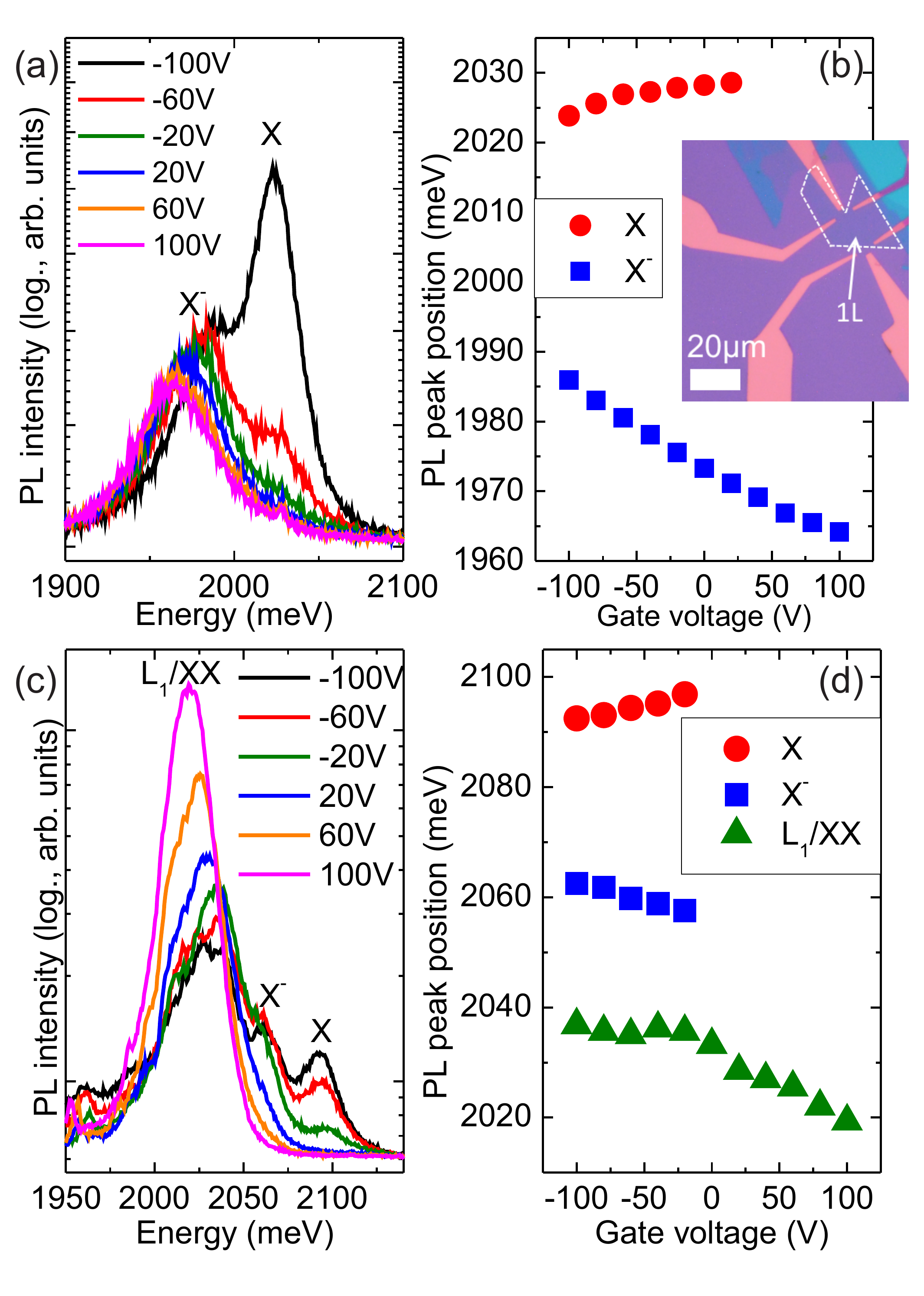}
\caption{(a) (a) PL spectra at room temperature for different gate voltages. (b) PL peak position of X and X$^-$ as a function of gate voltage. The inset shows an optical micrograph of the WS$_2$ flake on a Si/SiO$_2$ substrate with prestructured Ti:Au contacts. (c) PL spectra at $T=4$\,K for different gate voltages. (d) PL peak position of X, X$^-$ and L$_1$/XX peak as a function of gate voltage.}
\label{fig:Gate}
\end{figure}

Figure \ref{fig:TempSeries}(a) shows the PL spectra of single-layer WS$_2$ for various temperatures. In this measurement series, the laser excitation density is kept relatively low at 5~kWcm$^{-2}$ to avoid possible heating effects. At 295\,K, the spectrum consists of two peaks at 2018\,meV and 1975\,meV, which we attribute to the exciton (X) and the trion (X$^-$). The peak positions at room temperature are in very good agreement with recent reports~\cite{Cui15,Peim14,Bellus}. We note that even at room temperature, X and X$^-$ peaks can be separated due to their small linewidth. The existence of the trion peak indicates an intrinsic doping of our sample, as it is commonly observed also in other TMDCs~\cite{Kis_NatNano10}. When cooling down the sample, both PL peaks experience a blueshift in accordance with the Varshni equation~\cite{Varshni_Physica67}, which describes the change of the bandgap with temperature in a large variety of semiconductors:
\begin{equation}
E_g(T)=E_g(0)-\alpha T^2/(T+\beta),
\label{Varshni}
\end{equation}
where $E_g(0)$ is the bandgap at zero temperature and $\alpha$ and $\beta$ are phenomenological fit parameters. We assume that the exciton and trion binding energy are  temperature-independent, and that X and X$^-$ peaks rigidly shift with the bandgap.  We use   Eq.~\ref{Varshni} to fit the PL peak positions extracted for each temperature, as depicted in Fig.~\ref{fig:TempSeries}(b). For both peaks, the fit matches  with $\alpha=4.0\cdot 10^{-4}$\,eV/K, $\beta \approx 200$\,K and $E_g(0)=2088$\,meV for X and $E_g(0)=2045$\,meV for X$^-$. The parameters are comparable to those from previous studies on MoS$_2$ \cite{Korn11}. Our assignment of the X and X$^-$ peaks at $T=4$\,K is further confirmed by additional power-dependent PL measurements at 150~K~\cite{supp}, in which we observe a low-energy tail in the X$^-$ peak, which is typical of an electron-recoil effect and has recently also been observed for trions in MoSe$_2$ \cite{Ross13}.
The spectral weight shifts from X to X$^-$ with decreasing temperature. This indicates that the thermal energy at higher temperatures is large enough to lead to a partial dissociation of the trions. In Fig.~\ref{fig:TempSeries}(a), we also see that the X$^-$ peak develops a low-energy shoulder at 30\,K and, even more pronounced, at 4\,K, which we denominate as L$_1$/XX. We will demonstrate below that it stems from a superposition of defect-bound exciton (L$_1$) and biexciton (XX) emission.
In previous reports, either the L$_1$ and L$_2$ peak~\cite{Mitioglu} or the X$^-$ and L$_1$ peak~\cite{Hawrylak15,Zhang15},  have been attributed to exciton and trion emission. The actual X peak at about 2.09\,eV is absent in those studies. The fact that we see a well-pronounced exciton peak in our spectra might be due to our sample preparation process, which leads to a reduced interaction with the substrate in comparison to direct exfoliation of flakes onto SiO$_2$ using adhesive tape.

To confirm our assignment of the exciton and trion peaks, as well as the charge state of the trion, we perform gate-dependent PL measurements.  The inset in Fig.~\ref{fig:Gate}(b) shows a microscope image of a gated sample.  In Fig.~\ref{fig:Gate}(a),  PL spectra are plotted for different backgate voltages $V_{g}$ at room temperature. At large negative $V_g$, the X peak is the dominant one, whereas it is completely suppressed for positive $V_g$, where the X$^-$ peak is the only measurable feature. Hence, we infer that the trions in our samples are negatively charged.  This indicates that the WS$_2$ single-layer has a residual n-type doping, similar to MoS$_2$~\cite{Kis_NatNano10} but in contrast to WSe$_2$~\cite{Kim14}. Our room-temperature data is in perfect agreement with Ref.~\cite{Cui15}.
Figure~\ref{fig:Gate}(c) displays the gate-dependent PL spectra at 4\,K. For negative gate voltages, the X peak intensity  increases as the Fermi level is shifted towards the neutral regime. This clearly confirms the identification of the 2.088\,eV peak as the exciton peak. The X$^-$ peak, in contrast, gains in intensity  by increasing the gate voltage for $V_g > 0$.
In both gate-voltage dependent measurement series, we observe that the X$^-$ peak experiences a spectral redshift, while the X peak shows  a slight blueshift  with increasing $V_g$ (Fig. \ref{fig:Gate}(b) and (d)), so that the energy difference between X and X$^-$ peaks increases with increasing carrier concentration. This effect has also been observed in other TMDCs~\cite{Mak13,Ross13}. In the limit of low carrier concentration, the ionization energy of a trion is equal to the trion binding energy. In the presence of a 2D electron gas (2DEG), however, ionization of a trion requires that the ionized electron is excited to a state above the Fermi energy  of the 2DEG, as all states below the Fermi energy are occupied. Thus, the energy difference between  exciton and trion peaks is given by~\cite{Mak13}:
\begin{equation}
E_X - E_{X^-}=E_{b,X^-}+E_F,
\label{Fermi}
\end{equation}
with $E_X$ and $E_{X^-}$ being the exciton and trion PL peak energies, $E_{b,X^-}$ the trion binding energy and $E_F$ the Fermi energy, which is  proportional to $V_g$.  Due to intrinsic doping and the corresponding non-zero $E_F$, the measured exciton-trion energy difference $E_X-E_{X^-}$ of 43\,meV in the ungated sample shown in Fig.~\ref{fig:TempSeries}(a)  is larger than the actual trion binding energy. The exciton-trion peak separation in gated samples follows Eq.~\ref{Fermi}~\cite{supp}, showing a minimal peak separation of 30\,meV at $V_g=-100$\,V. This represents an upper limit for the trion binding energy.
\begin{figure}[htb]%
\includegraphics*[width=\linewidth]{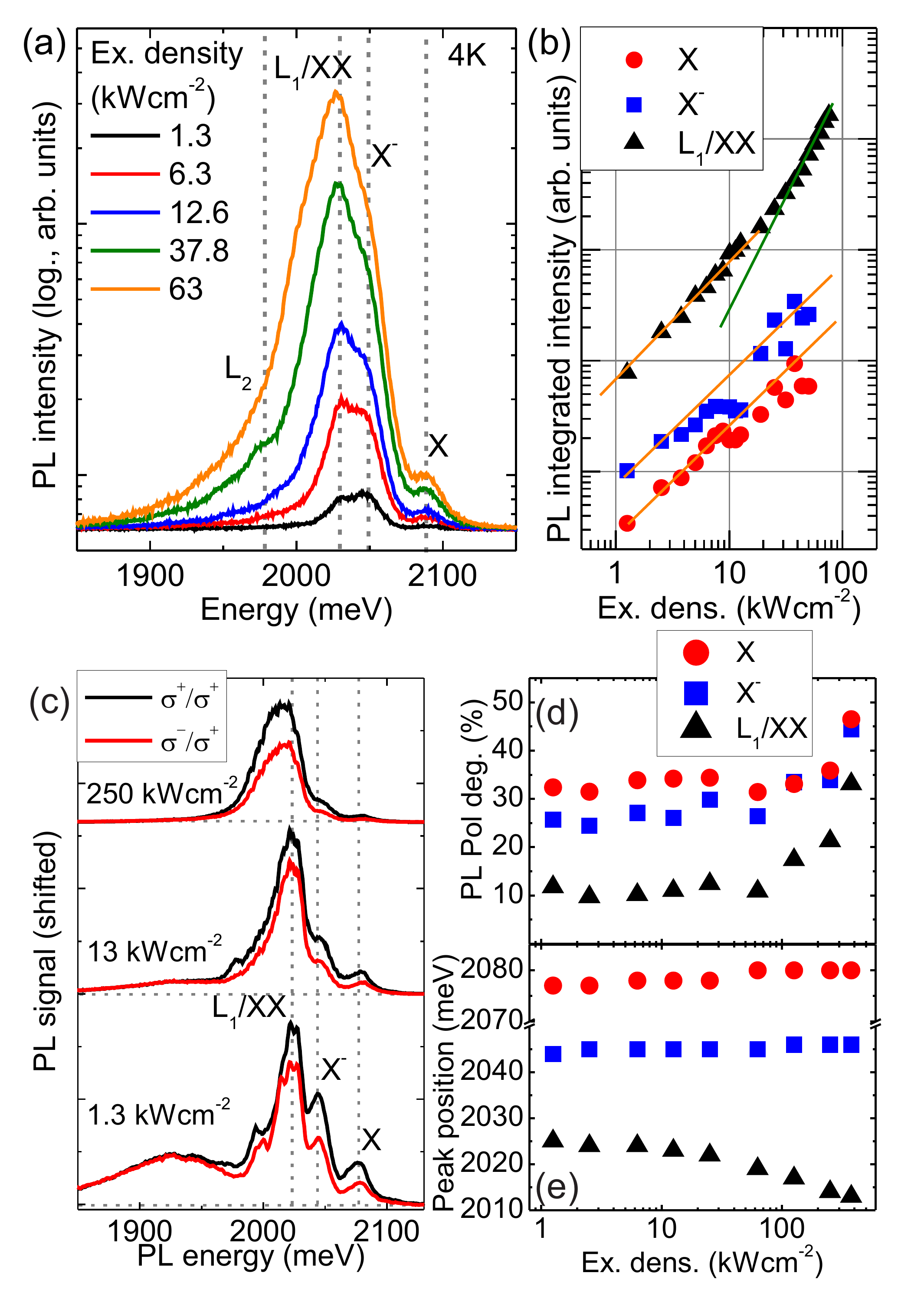}
\caption{(a) PL spectra of single-layer WS$_2$ at $T=4$\,K for various excitation densities. (b) Double-logarithmic plot of integrated PL intensity of X (red circles), X$^-$ (blue squares) and L$_1$/XX peak (black triangles) as a function of excitation density. The orange solid lines indicate a linear dependency, whereas the green solid line indicates a quadratic dependency. (c) Helicity-resolved PL spectra of single-layer WS$_2$ at $T=4$\,K under near-resonant, circularly-polarized excitation for various excitation densities. The black and red spectra are for co-circular and contra-circular excitation and detection, respectively. (d) PL circular polarization degree and (e) PL peak position for X, X$^-$ and L$_1$/XX peaks as a function of excitation density.}
\label{fig:PowerSeries}
\end{figure}

Finally, we focus on the low-energy feature labeled as L$_1$/XX that arises at temperatures below 60\,K. Fig.~\ref{fig:PowerSeries}(a) shows the PL spectra at $T=4$\,K for different excitation powers. Whereas at low powers, X$^-$ and L$_1$/XX are spectrally well separated and of similar intensity, at higher excitation powers, the  L$_1$/XX peak completely dominates the spectrum. Additionally, a second low-energy peak L$_2$ with moderate intensity is discernible around 1.98\,eV. It may stem from defect-bound excitons, as its intensity decreases relative to the other peaks with increasing excitation density. To get a better insight into the nature of the L$_1$/XX peak, we extract the integrated PL intensity for L$_1$/XX, X$^-$ and X for different excitation densities, as displayed in the double-logarithmic graph in Fig.~\ref{fig:PowerSeries}(b). X and X$^-$ show a rather linear behavior indicated by the orange solid line, as expected for an excitonic feature \cite{Kling07}. In contrast, the L$_1$/XX peak  exhibits a linear dependence at low excitation density, while for excitation densities larger than 25~kWcm$^{-2}$, the data is well-described by a quadratic fit, indicated by the green solid line in Fig.~\ref{fig:PowerSeries}(b). Such a quadratic increase in PL emission intensity  is  expected for biexcitons~\cite{Gourley}, although smaller, superlinear slopes are often observed in experiment due to the kinetics of biexciton formation and exciton recombination~\cite{Heinz15}. The  different behavior for low and high excitation densities indicates that in fact, two different emission lines are responsible for the observed L$_1$/XX peak: at low excitation density, the main contribution to the PL at the L$_1$/XX peak position stems from defect-bound excitons (denominated  L$_1$). At high excitation density, the biexciton (XX) emission is dominant. To confirm our interpretation, we perform an excitation-density dependent measurement series utilizing near-resonant, circularly-polarized excitation. Figure~\ref{fig:PowerSeries}(c) shows helicity-resolved PL spectra measured at 4\,K using different excitation densities. At low excitation density, L$_1$/XX, X$^-$ and X peaks are clearly observable, together with a spectrally broad feature at lower energy. This feature is reminiscent of low-temperature PL spectra of MoS$_2$, where it is attributed to surface-adsorbate-bound excitons~\cite{Plechinger12}. For this feature, co- and contra-circularly-polarized PL spectra have the same intensity, indicating no circular polarization. By contrast, L$_1$/XX, X$^-$ and X peaks show a clear intensity difference in the helicity-resolved PL. With increasing excitation density, the L$_1$/XX  emission begins to dominate the spectrum.  For higher excitation densities, in the same range for which we observe the quadratic increase of the PL intensity discussed above, the L$_1$/XX peak position shows a pronounced redshift and its polarization degree increases. These two observations are analyzed and compared to the behavior of the X$^-$ and X peaks in Fig.~\ref{fig:PowerSeries}(d) and (e). The circular polarization degree of the PL emission in single-layer TMDCs is an indicator of  valley polarization, and for defect-related PL peaks, low values have been reported. By contrast, excitons, trions and biexcitons should  show a significant PL polarization degree under near-resonant excitation~\cite{Heinz15}. In Fig.~\ref{fig:PowerSeries}(d), we show that the PL polarization for the X$^-$ and X peaks is high and remains almost constant throughout the investigated excitation density range. By contrast, the L$_1$/XX peak has a low PL polarization degree at low excitation density, indicative of defect-related PL emission. The PL polarization degree increases with increasing excitation density, as expected for biexciton emission, reaching similar values as the X$^-$ peak for the highest excitation density values in our series. As shown in Fig.~\ref{fig:PowerSeries}(e), the L$_1$/XX peak redshifts by about 10~meV in the investigated excitation density range. This indicates that the L$_1$ emission from defect-bound excitons at low excitation density is at a higher energy than the biexciton emission at high excitation density. We  exclude  local heating induced by the laser as a source of the redshift for the L$_1$/XX, since  neither X$^-$ or X peaks display a redshift -- by contrast, they show a slight blueshift.  Thus, we can interpret the energy separation of about 65\,meV between the X and XX features as the biexciton binding energy $E_{b,XX}$. Currently, the value of the exciton binding energy in single-layer WS$_2$ is still under discussion. The values determined in different experiments range between 320~meV~\cite{Chernikov_Rydberg_PRL14} and 700~meV~\cite{Zhang14}. Thus, the Haynes factor, i.e., the ratio of $E_{b,XX}$ and the exciton binding energy,  ranges between 9 and 20~percent, which is comparable to values for biexcitons in quantum wells~\cite{Kling07} and those observed in WSe$_2$~\cite{Heinz15}. Remarkably, in our WS$_2$ samples, strong biexciton PL emission is observable already under cw laser excitation, while pulsed excitation was required to study biexciton emission in WSe$_2$~\cite{Heinz15}. This indicates pronounced differences in the kinetics of biexciton formation in different TMDCs.

In conclusion, we have presented temperature-dependent PL measurements on mechanically exfoliated single-layer WS$_2$. We find that the exciton and trion peaks are well separated even in the room temperature spectrum and their emission can be tracked down to 4\,K. By tuning the Fermi level in our samples, we can unambiguously assign the 2.09\,eV PL peak to exciton and the 2.05\,eV PL peak to trion emission at $T=4$\,K. At low temperatures, we observe the emergence of a lower-energy peak, which we identify as a superposition of defect-bound exciton and biexciton emission by the power dependence of its emission intensity and circular polarization degree.  These results  clarify some issues in the interpretation of low-temperature PL spectra in single-layer WS$_2$, which is a promising candidate for all-2D electrooptical and valleytronic devices.
\section*{Acknowledgements}
The authors ackowledge financial support by the DFG via KO3612/1-1, GRK1570 and SFB689.
\bibliography{WS2}

\begin{thebibliography}{32}%
\makeatletter
\providecommand \@ifxundefined [1]{%
 \@ifx{#1\undefined}
}%
\providecommand \@ifnum [1]{%
 \ifnum #1\expandafter \@firstoftwo
 \else \expandafter \@secondoftwo
 \fi
}%
\providecommand \@ifx [1]{%
 \ifx #1\expandafter \@firstoftwo
 \else \expandafter \@secondoftwo
 \fi
}%
\providecommand \natexlab [1]{#1}%
\providecommand \enquote  [1]{``#1''}%
\providecommand \bibnamefont  [1]{#1}%
\providecommand \bibfnamefont [1]{#1}%
\providecommand \citenamefont [1]{#1}%
\providecommand \href@noop [0]{\@secondoftwo}%
\providecommand \href [0]{\begingroup \@sanitize@url \@href}%
\providecommand \@href[1]{\@@startlink{#1}\@@href}%
\providecommand \@@href[1]{\endgroup#1\@@endlink}%
\providecommand \@sanitize@url [0]{\catcode `\\12\catcode `\$12\catcode
  `\&12\catcode `\#12\catcode `\^12\catcode `\_12\catcode `\%12\relax}%
\providecommand \@@startlink[1]{}%
\providecommand \@@endlink[0]{}%
\providecommand \url  [0]{\begingroup\@sanitize@url \@url }%
\providecommand \@url [1]{\endgroup\@href {#1}{\urlprefix }}%
\providecommand \urlprefix  [0]{URL }%
\providecommand \Eprint [0]{\href }%
\providecommand \doibase [0]{http://dx.doi.org/}%
\providecommand \selectlanguage [0]{\@gobble}%
\providecommand \bibinfo  [0]{\@secondoftwo}%
\providecommand \bibfield  [0]{\@secondoftwo}%
\providecommand \translation [1]{[#1]}%
\providecommand \BibitemOpen [0]{}%
\providecommand \bibitemStop [0]{}%
\providecommand \bibitemNoStop [0]{.\EOS\space}%
\providecommand \EOS [0]{\spacefactor3000\relax}%
\providecommand \BibitemShut  [1]{\csname bibitem#1\endcsname}%
\let\auto@bib@innerbib\@empty
\bibitem [{\citenamefont {Mak}\ \emph {et~al.}(2010)\citenamefont {Mak},
  \citenamefont {Lee}, \citenamefont {Hone}, \citenamefont {Shan},\ and\
  \citenamefont {Heinz}}]{HeinzPRL10}%
  \BibitemOpen
  \bibfield  {author} {\bibinfo {author} {\bibfnamefont {K.~F.}\ \bibnamefont
  {Mak}}, \bibinfo {author} {\bibfnamefont {C.}~\bibnamefont {Lee}}, \bibinfo
  {author} {\bibfnamefont {J.}~\bibnamefont {Hone}}, \bibinfo {author}
  {\bibfnamefont {J.}~\bibnamefont {Shan}}, \ and\ \bibinfo {author}
  {\bibfnamefont {T.~F.}\ \bibnamefont {Heinz}},\ }\href {\doibase
  10.1103/PhysRevLett.105.136805} {\bibfield  {journal} {\bibinfo  {journal}
  {Phys. Rev. Lett.}\ }\textbf {\bibinfo {volume} {105}},\ \bibinfo {pages}
  {136805} (\bibinfo {year} {2010})}\BibitemShut {NoStop}%
\bibitem [{\citenamefont {Radisavljevic}\ \emph {et~al.}(2011)\citenamefont
  {Radisavljevic}, \citenamefont {Radenovic}, \citenamefont {Brivio},
  \citenamefont {Giacometti},\ and\ \citenamefont {Kis}}]{Kis_NatNano10}%
  \BibitemOpen
  \bibfield  {author} {\bibinfo {author} {\bibfnamefont {B.}~\bibnamefont
  {Radisavljevic}}, \bibinfo {author} {\bibfnamefont {A.}~\bibnamefont
  {Radenovic}}, \bibinfo {author} {\bibfnamefont {J.}~\bibnamefont {Brivio}},
  \bibinfo {author} {\bibfnamefont {V.}~\bibnamefont {Giacometti}}, \ and\
  \bibinfo {author} {\bibfnamefont {A.}~\bibnamefont {Kis}},\ }\href@noop {}
  {\bibfield  {journal} {\bibinfo  {journal} {Nat. Nanotechnol.}\ }\textbf
  {\bibinfo {volume} {6}},\ \bibinfo {pages} {147} (\bibinfo {year}
  {2011})}\BibitemShut {NoStop}%
\bibitem [{\citenamefont {Mak}\ \emph {et~al.}(2012)\citenamefont {Mak},
  \citenamefont {He}, \citenamefont {Shan},\ and\ \citenamefont
  {Heinz}}]{Mak12}%
  \BibitemOpen
  \bibfield  {author} {\bibinfo {author} {\bibfnamefont {K.~F.}\ \bibnamefont
  {Mak}}, \bibinfo {author} {\bibfnamefont {K.}~\bibnamefont {He}}, \bibinfo
  {author} {\bibfnamefont {J.}~\bibnamefont {Shan}}, \ and\ \bibinfo {author}
  {\bibfnamefont {T.~F.}\ \bibnamefont {Heinz}},\ }\href@noop {} {\bibfield
  {journal} {\bibinfo  {journal} {Nat. Nanotechnol.}\ }\textbf {\bibinfo
  {volume} {7}},\ \bibinfo {pages} {494} (\bibinfo {year} {2012})}\BibitemShut
  {NoStop}%
\bibitem [{\citenamefont {Furchi}\ \emph {et~al.}(2014)\citenamefont {Furchi},
  \citenamefont {Pospischil}, \citenamefont {Libisch}, \citenamefont
  {Burgd\"orfer},\ and\ \citenamefont {Mueller}}]{Mueller14}%
  \BibitemOpen
  \bibfield  {author} {\bibinfo {author} {\bibfnamefont {M.~M.}\ \bibnamefont
  {Furchi}}, \bibinfo {author} {\bibfnamefont {A.}~\bibnamefont {Pospischil}},
  \bibinfo {author} {\bibfnamefont {F.}~\bibnamefont {Libisch}}, \bibinfo
  {author} {\bibfnamefont {J.}~\bibnamefont {Burgd\"orfer}}, \ and\ \bibinfo
  {author} {\bibfnamefont {T.}~\bibnamefont {Mueller}},\ }\href@noop {}
  {\bibfield  {journal} {\bibinfo  {journal} {Nano Lett.}\ }\textbf {\bibinfo
  {volume} {14}},\ \bibinfo {pages} {4785} (\bibinfo {year}
  {2014})}\BibitemShut {NoStop}%
\bibitem [{\citenamefont {Splendiani}\ \emph {et~al.}(2010)\citenamefont
  {Splendiani}, \citenamefont {Sun}, \citenamefont {Zhang}, \citenamefont {Li},
  \citenamefont {Kim}, \citenamefont {Chim}, \citenamefont {Galli},\ and\
  \citenamefont {Wang}}]{Splen_Nano10}%
  \BibitemOpen
  \bibfield  {author} {\bibinfo {author} {\bibfnamefont {A.}~\bibnamefont
  {Splendiani}}, \bibinfo {author} {\bibfnamefont {L.}~\bibnamefont {Sun}},
  \bibinfo {author} {\bibfnamefont {Y.}~\bibnamefont {Zhang}}, \bibinfo
  {author} {\bibfnamefont {T.}~\bibnamefont {Li}}, \bibinfo {author}
  {\bibfnamefont {J.}~\bibnamefont {Kim}}, \bibinfo {author} {\bibfnamefont
  {C.-Y.}\ \bibnamefont {Chim}}, \bibinfo {author} {\bibfnamefont
  {G.}~\bibnamefont {Galli}}, \ and\ \bibinfo {author} {\bibfnamefont
  {F.}~\bibnamefont {Wang}},\ }\href {\doibase 10.1021/nl903868w} {\bibfield
  {journal} {\bibinfo  {journal} {Nano Lett.}\ }\textbf {\bibinfo {volume}
  {10}},\ \bibinfo {pages} {1271} (\bibinfo {year} {2010})}\BibitemShut
  {NoStop}%
\bibitem [{\citenamefont {Cheiwchanchamnangij}\ and\ \citenamefont
  {Lambrecht}(2012)}]{Chei12}%
  \BibitemOpen
  \bibfield  {author} {\bibinfo {author} {\bibfnamefont {T.}~\bibnamefont
  {Cheiwchanchamnangij}}\ and\ \bibinfo {author} {\bibfnamefont {W.~R.~L.}\
  \bibnamefont {Lambrecht}},\ }\href@noop {} {\bibfield  {journal} {\bibinfo
  {journal} {Phys. Rev. B}\ }\textbf {\bibinfo {volume} {85}},\ \bibinfo
  {pages} {205302} (\bibinfo {year} {2012})}\BibitemShut {NoStop}%
\bibitem [{\citenamefont {Bergh\"auser}\ and\ \citenamefont
  {Malic}(2014)}]{Malic14}%
  \BibitemOpen
  \bibfield  {author} {\bibinfo {author} {\bibfnamefont {G.}~\bibnamefont
  {Bergh\"auser}}\ and\ \bibinfo {author} {\bibfnamefont {E.}~\bibnamefont
  {Malic}},\ }\href@noop {} {\bibfield  {journal} {\bibinfo  {journal} {Phys.
  Rev. B}\ }\textbf {\bibinfo {volume} {89}},\ \bibinfo {pages} {125309}
  (\bibinfo {year} {2014})}\BibitemShut {NoStop}%
\bibitem [{\citenamefont {Chernikov}\ \emph {et~al.}(2014)\citenamefont
  {Chernikov}, \citenamefont {Berkelbach}, \citenamefont {Hill}, \citenamefont
  {Rigosi}, \citenamefont {Li}, \citenamefont {Aslan}, \citenamefont
  {Reichman}, \citenamefont {Hybertsen},\ and\ \citenamefont
  {Heinz}}]{Chernikov_Rydberg_PRL14}%
  \BibitemOpen
  \bibfield  {author} {\bibinfo {author} {\bibfnamefont {A.}~\bibnamefont
  {Chernikov}}, \bibinfo {author} {\bibfnamefont {T.~C.}\ \bibnamefont
  {Berkelbach}}, \bibinfo {author} {\bibfnamefont {H.~M.}\ \bibnamefont
  {Hill}}, \bibinfo {author} {\bibfnamefont {A.}~\bibnamefont {Rigosi}},
  \bibinfo {author} {\bibfnamefont {Y.}~\bibnamefont {Li}}, \bibinfo {author}
  {\bibfnamefont {O.~B.}\ \bibnamefont {Aslan}}, \bibinfo {author}
  {\bibfnamefont {D.~R.}\ \bibnamefont {Reichman}}, \bibinfo {author}
  {\bibfnamefont {M.~S.}\ \bibnamefont {Hybertsen}}, \ and\ \bibinfo {author}
  {\bibfnamefont {T.~F.}\ \bibnamefont {Heinz}},\ }\href {\doibase
  10.1103/PhysRevLett.113.076802} {\bibfield  {journal} {\bibinfo  {journal}
  {Phys. Rev. Lett.}\ }\textbf {\bibinfo {volume} {113}},\ \bibinfo {pages}
  {076802} (\bibinfo {year} {2014})}\BibitemShut {NoStop}%
\bibitem [{\citenamefont {Ye}\ \emph {et~al.}(2014)\citenamefont {Ye},
  \citenamefont {Cao}, \citenamefont {O'Brien}, \citenamefont {Zhu},
  \citenamefont {Yin}, \citenamefont {Wang}, \citenamefont {Louie},\ and\
  \citenamefont {Zhang}}]{Zhang14}%
  \BibitemOpen
  \bibfield  {author} {\bibinfo {author} {\bibfnamefont {Z.}~\bibnamefont
  {Ye}}, \bibinfo {author} {\bibfnamefont {T.}~\bibnamefont {Cao}}, \bibinfo
  {author} {\bibfnamefont {K.}~\bibnamefont {O'Brien}}, \bibinfo {author}
  {\bibfnamefont {H.}~\bibnamefont {Zhu}}, \bibinfo {author} {\bibfnamefont
  {X.}~\bibnamefont {Yin}}, \bibinfo {author} {\bibfnamefont {Y.}~\bibnamefont
  {Wang}}, \bibinfo {author} {\bibfnamefont {S.~G.}\ \bibnamefont {Louie}}, \
  and\ \bibinfo {author} {\bibfnamefont {X.}~\bibnamefont {Zhang}},\
  }\href@noop {} {\bibfield  {journal} {\bibinfo  {journal} {Nature}\ }\textbf
  {\bibinfo {volume} {513}},\ \bibinfo {pages} {214} (\bibinfo {year}
  {2014})}\BibitemShut {NoStop}%
\bibitem [{\citenamefont {Zhu}\ \emph {et~al.}(2015)\citenamefont {Zhu},
  \citenamefont {Chen},\ and\ \citenamefont {Cui}}]{Cui15}%
  \BibitemOpen
  \bibfield  {author} {\bibinfo {author} {\bibfnamefont {B.}~\bibnamefont
  {Zhu}}, \bibinfo {author} {\bibfnamefont {X.}~\bibnamefont {Chen}}, \ and\
  \bibinfo {author} {\bibfnamefont {X.}~\bibnamefont {Cui}},\ }\href {\doibase
  10.1038/srep09218} {\bibfield  {journal} {\bibinfo  {journal} {Scientific
  Reports}\ }\textbf {\bibinfo {volume} {5}},\ \bibinfo {pages} {9218}
  (\bibinfo {year} {2015})}\BibitemShut {NoStop}%
\bibitem [{\citenamefont {Hanbicki}\ \emph {et~al.}(2015)\citenamefont
  {Hanbicki}, \citenamefont {Currie}, \citenamefont {Kioseoglou}, \citenamefont
  {Friedman},\ and\ \citenamefont {Jonker}}]{Jonker15}%
  \BibitemOpen
  \bibfield  {author} {\bibinfo {author} {\bibfnamefont {A.~T.}\ \bibnamefont
  {Hanbicki}}, \bibinfo {author} {\bibfnamefont {M.}~\bibnamefont {Currie}},
  \bibinfo {author} {\bibfnamefont {G.}~\bibnamefont {Kioseoglou}}, \bibinfo
  {author} {\bibfnamefont {A.~L.}\ \bibnamefont {Friedman}}, \ and\ \bibinfo
  {author} {\bibfnamefont {B.~T.}\ \bibnamefont {Jonker}},\ }\href@noop {}
  {\bibfield  {journal} {\bibinfo  {journal} {Solid State Comm.}\ }\textbf
  {\bibinfo {volume} {203}},\ \bibinfo {pages} {16} (\bibinfo {year}
  {2015})}\BibitemShut {NoStop}%
\bibitem [{\citenamefont {Mak}\ \emph {et~al.}(2013)\citenamefont {Mak},
  \citenamefont {He}, \citenamefont {Lee}, \citenamefont {Lee}, \citenamefont
  {Hone}, \citenamefont {Heinz},\ and\ \citenamefont {Shan}}]{Mak13}%
  \BibitemOpen
  \bibfield  {author} {\bibinfo {author} {\bibfnamefont {K.~F.}\ \bibnamefont
  {Mak}}, \bibinfo {author} {\bibfnamefont {K.}~\bibnamefont {He}}, \bibinfo
  {author} {\bibfnamefont {C.}~\bibnamefont {Lee}}, \bibinfo {author}
  {\bibfnamefont {G.~H.}\ \bibnamefont {Lee}}, \bibinfo {author} {\bibfnamefont
  {J.}~\bibnamefont {Hone}}, \bibinfo {author} {\bibfnamefont {T.~F.}\
  \bibnamefont {Heinz}}, \ and\ \bibinfo {author} {\bibfnamefont
  {J.}~\bibnamefont {Shan}},\ }\href@noop {} {\bibfield  {journal} {\bibinfo
  {journal} {Nat. Mater.}\ }\textbf {\bibinfo {volume} {12}},\ \bibinfo {pages}
  {207} (\bibinfo {year} {2013})}\BibitemShut {NoStop}%
\bibitem [{\citenamefont {Ross}\ \emph {et~al.}(2013)\citenamefont {Ross},
  \citenamefont {Wu}, \citenamefont {Yu}, \citenamefont {Ghimire},
  \citenamefont {Jones}, \citenamefont {Aivazian}, \citenamefont {Yan},
  \citenamefont {Mandrus}, \citenamefont {Xiao}, \citenamefont {Yao},\ and\
  \citenamefont {Xu}}]{Ross13}%
  \BibitemOpen
  \bibfield  {author} {\bibinfo {author} {\bibfnamefont {J.~S.}\ \bibnamefont
  {Ross}}, \bibinfo {author} {\bibfnamefont {S.~F.}\ \bibnamefont {Wu}},
  \bibinfo {author} {\bibfnamefont {H.~Y.}\ \bibnamefont {Yu}}, \bibinfo
  {author} {\bibfnamefont {N.~J.}\ \bibnamefont {Ghimire}}, \bibinfo {author}
  {\bibfnamefont {A.~M.}\ \bibnamefont {Jones}}, \bibinfo {author}
  {\bibfnamefont {G.}~\bibnamefont {Aivazian}}, \bibinfo {author}
  {\bibfnamefont {J.~Q.}\ \bibnamefont {Yan}}, \bibinfo {author} {\bibfnamefont
  {D.~G.}\ \bibnamefont {Mandrus}}, \bibinfo {author} {\bibfnamefont
  {D.}~\bibnamefont {Xiao}}, \bibinfo {author} {\bibfnamefont {W.}~\bibnamefont
  {Yao}}, \ and\ \bibinfo {author} {\bibfnamefont {X.}~\bibnamefont {Xu}},\
  }\href@noop {} {\bibfield  {journal} {\bibinfo  {journal} {Nat. Commun.}\
  }\textbf {\bibinfo {volume} {4}},\ \bibinfo {pages} {1474} (\bibinfo {year}
  {2013})}\BibitemShut {NoStop}%
\bibitem [{\citenamefont {Jones}\ \emph {et~al.}(2013)\citenamefont {Jones},
  \citenamefont {Yu}, \citenamefont {Ghimire}, \citenamefont {Wu},
  \citenamefont {Aivazian}, \citenamefont {Ross}, \citenamefont {Zhao},
  \citenamefont {Yan}, \citenamefont {Mandrus}, \citenamefont {Xiao},
  \citenamefont {Yao},\ and\ \citenamefont {Xu}}]{Jones13}%
  \BibitemOpen
  \bibfield  {author} {\bibinfo {author} {\bibfnamefont {A.~M.}\ \bibnamefont
  {Jones}}, \bibinfo {author} {\bibfnamefont {H.}~\bibnamefont {Yu}}, \bibinfo
  {author} {\bibfnamefont {N.~J.}\ \bibnamefont {Ghimire}}, \bibinfo {author}
  {\bibfnamefont {S.}~\bibnamefont {Wu}}, \bibinfo {author} {\bibfnamefont
  {G.}~\bibnamefont {Aivazian}}, \bibinfo {author} {\bibfnamefont {J.~S.}\
  \bibnamefont {Ross}}, \bibinfo {author} {\bibfnamefont {B.}~\bibnamefont
  {Zhao}}, \bibinfo {author} {\bibfnamefont {J.}~\bibnamefont {Yan}}, \bibinfo
  {author} {\bibfnamefont {D.~G.}\ \bibnamefont {Mandrus}}, \bibinfo {author}
  {\bibfnamefont {D.}~\bibnamefont {Xiao}}, \bibinfo {author} {\bibfnamefont
  {W.}~\bibnamefont {Yao}}, \ and\ \bibinfo {author} {\bibfnamefont
  {X.}~\bibnamefont {Xu}},\ }\href@noop {} {\bibfield  {journal} {\bibinfo
  {journal} {Nat. Nanotechnol.}\ }\textbf {\bibinfo {volume} {8}},\ \bibinfo
  {pages} {634} (\bibinfo {year} {2013})}\BibitemShut {NoStop}%
\bibitem [{\citenamefont {Plechinger}\ \emph {et~al.}(2012)\citenamefont
  {Plechinger}, \citenamefont {Schrettenbrunner}, \citenamefont {Eroms},
  \citenamefont {Weiss}, \citenamefont {Sch\"{u}ller},\ and\ \citenamefont
  {Korn}}]{Plechinger12}%
  \BibitemOpen
  \bibfield  {author} {\bibinfo {author} {\bibfnamefont {G.}~\bibnamefont
  {Plechinger}}, \bibinfo {author} {\bibfnamefont {F.-X.}\ \bibnamefont
  {Schrettenbrunner}}, \bibinfo {author} {\bibfnamefont {J.}~\bibnamefont
  {Eroms}}, \bibinfo {author} {\bibfnamefont {D.}~\bibnamefont {Weiss}},
  \bibinfo {author} {\bibfnamefont {C.}~\bibnamefont {Sch\"{u}ller}}, \ and\
  \bibinfo {author} {\bibfnamefont {T.}~\bibnamefont {Korn}},\ }\href@noop {}
  {\bibfield  {journal} {\bibinfo  {journal} {physica status solidi (RRL) –
  Rapid Research Letters}\ }\textbf {\bibinfo {volume} {6}},\ \bibinfo {pages}
  {126} (\bibinfo {year} {2012})}\BibitemShut {NoStop}%
\bibitem [{\citenamefont {Wang}\ \emph
  {et~al.}(2014{\natexlab{a}})\citenamefont {Wang}, \citenamefont {Bouet},
  \citenamefont {Lagarde}, \citenamefont {Vidal}, \citenamefont {Balocchi},
  \citenamefont {Amand}, \citenamefont {Marie},\ and\ \citenamefont
  {Urbaszek}}]{Wang14}%
  \BibitemOpen
  \bibfield  {author} {\bibinfo {author} {\bibfnamefont {G.}~\bibnamefont
  {Wang}}, \bibinfo {author} {\bibfnamefont {L.}~\bibnamefont {Bouet}},
  \bibinfo {author} {\bibfnamefont {D.}~\bibnamefont {Lagarde}}, \bibinfo
  {author} {\bibfnamefont {M.}~\bibnamefont {Vidal}}, \bibinfo {author}
  {\bibfnamefont {A.}~\bibnamefont {Balocchi}}, \bibinfo {author}
  {\bibfnamefont {T.}~\bibnamefont {Amand}}, \bibinfo {author} {\bibfnamefont
  {X.}~\bibnamefont {Marie}}, \ and\ \bibinfo {author} {\bibfnamefont
  {B.}~\bibnamefont {Urbaszek}},\ }\href@noop {} {\bibfield  {journal}
  {\bibinfo  {journal} {Phys. Rev. B}\ }\textbf {\bibinfo {volume} {90}},\
  \bibinfo {pages} {075413} (\bibinfo {year} {2014}{\natexlab{a}})}\BibitemShut
  {NoStop}%
\bibitem [{\citenamefont {Wang}\ \emph
  {et~al.}(2014{\natexlab{b}})\citenamefont {Wang}, \citenamefont {Bouet},
  \citenamefont {Lagarde}, \citenamefont {Vidal}, \citenamefont {Balocchi},
  \citenamefont {Amand}, \citenamefont {Marie},\ and\ \citenamefont
  {Urbaszek}}]{Marie14}%
  \BibitemOpen
  \bibfield  {author} {\bibinfo {author} {\bibfnamefont {G.}~\bibnamefont
  {Wang}}, \bibinfo {author} {\bibfnamefont {L.}~\bibnamefont {Bouet}},
  \bibinfo {author} {\bibfnamefont {D.}~\bibnamefont {Lagarde}}, \bibinfo
  {author} {\bibfnamefont {M.}~\bibnamefont {Vidal}}, \bibinfo {author}
  {\bibfnamefont {A.}~\bibnamefont {Balocchi}}, \bibinfo {author}
  {\bibfnamefont {T.}~\bibnamefont {Amand}}, \bibinfo {author} {\bibfnamefont
  {X.}~\bibnamefont {Marie}}, \ and\ \bibinfo {author} {\bibfnamefont
  {B.}~\bibnamefont {Urbaszek}},\ }\href {\doibase 10.1103/PhysRevB.90.075413}
  {\bibfield  {journal} {\bibinfo  {journal} {Phys. Rev. B}\ }\textbf {\bibinfo
  {volume} {90}},\ \bibinfo {pages} {075413} (\bibinfo {year}
  {2014}{\natexlab{b}})}\BibitemShut {NoStop}%
\bibitem [{\citenamefont {Klingshirn}(2007)}]{Kling07}%
  \BibitemOpen
  \bibinfo {editor} {\bibfnamefont {C.~F.}\ \bibnamefont {Klingshirn}},\ ed.,\
  \href@noop {} {\emph {\bibinfo {title} {Semiconductor Optics}}},\
  Vol.~\bibinfo {volume} {3}\ (\bibinfo  {publisher} {Springer},\ \bibinfo
  {year} {2007})\BibitemShut {NoStop}%
\bibitem [{\citenamefont {You}\ \emph {et~al.}(2015)\citenamefont {You},
  \citenamefont {Zhang}, \citenamefont {Berkelbach}, \citenamefont {Hybertsen},
  \citenamefont {Reichman},\ and\ \citenamefont {Heinz}}]{Heinz15}%
  \BibitemOpen
  \bibfield  {author} {\bibinfo {author} {\bibfnamefont {Y.}~\bibnamefont
  {You}}, \bibinfo {author} {\bibfnamefont {X.-X.}\ \bibnamefont {Zhang}},
  \bibinfo {author} {\bibfnamefont {T.~C.}\ \bibnamefont {Berkelbach}},
  \bibinfo {author} {\bibfnamefont {M.~S.}\ \bibnamefont {Hybertsen}}, \bibinfo
  {author} {\bibfnamefont {D.~R.}\ \bibnamefont {Reichman}}, \ and\ \bibinfo
  {author} {\bibfnamefont {T.~F.}\ \bibnamefont {Heinz}},\ }\href@noop {}
  {\bibfield  {journal} {\bibinfo  {journal} {Nat. Phys.}\ }\textbf {\bibinfo
  {volume} {11}},\ \bibinfo {pages} {477} (\bibinfo {year} {2015})}\BibitemShut
  {NoStop}%
\bibitem [{\citenamefont {Scrace}\ \emph {et~al.}(2015)\citenamefont {Scrace},
  \citenamefont {Tsai}, \citenamefont {Barman}, \citenamefont {Scheidenback},
  \citenamefont {Petrou}, \citenamefont {Kioseoglou}, \citenamefont {Ozfidan},
  \citenamefont {Korkusinski},\ and\ \citenamefont {Hawrylak}}]{Hawrylak15}%
  \BibitemOpen
  \bibfield  {author} {\bibinfo {author} {\bibfnamefont {T.}~\bibnamefont
  {Scrace}}, \bibinfo {author} {\bibfnamefont {Y.}~\bibnamefont {Tsai}},
  \bibinfo {author} {\bibfnamefont {B.}~\bibnamefont {Barman}}, \bibinfo
  {author} {\bibfnamefont {L.}~\bibnamefont {Scheidenback}}, \bibinfo {author}
  {\bibfnamefont {A.}~\bibnamefont {Petrou}}, \bibinfo {author} {\bibfnamefont
  {G.}~\bibnamefont {Kioseoglou}}, \bibinfo {author} {\bibfnamefont
  {I.}~\bibnamefont {Ozfidan}}, \bibinfo {author} {\bibfnamefont
  {M.}~\bibnamefont {Korkusinski}}, \ and\ \bibinfo {author} {\bibfnamefont
  {P.}~\bibnamefont {Hawrylak}},\ }\href {\doibase 10.1038/nnano.2015.78}
  {\bibfield  {journal} {\bibinfo  {journal} {Nat. Nanotechnol.}\ } (\bibinfo
  {year} {2015}),\ 10.1038/nnano.2015.78},\ \bibinfo {note}
  {2015/05/11/online}\BibitemShut {NoStop}%
\bibitem [{\citenamefont {Mitioglu}\ \emph {et~al.}(2013)\citenamefont
  {Mitioglu}, \citenamefont {Plochocka}, \citenamefont {Jadczak}, \citenamefont
  {Escoffier}, \citenamefont {Rikken}, \citenamefont {Kulyuk},\ and\
  \citenamefont {Maude}}]{Mitioglu}%
  \BibitemOpen
  \bibfield  {author} {\bibinfo {author} {\bibfnamefont {A.~A.}\ \bibnamefont
  {Mitioglu}}, \bibinfo {author} {\bibfnamefont {P.}~\bibnamefont {Plochocka}},
  \bibinfo {author} {\bibfnamefont {J.~N.}\ \bibnamefont {Jadczak}}, \bibinfo
  {author} {\bibfnamefont {W.}~\bibnamefont {Escoffier}}, \bibinfo {author}
  {\bibfnamefont {G.~L. J.~A.}\ \bibnamefont {Rikken}}, \bibinfo {author}
  {\bibfnamefont {L.}~\bibnamefont {Kulyuk}}, \ and\ \bibinfo {author}
  {\bibfnamefont {D.~K.}\ \bibnamefont {Maude}},\ }\href {\doibase
  http://dx.doi.org/10.1103/PhysRevB.88.245403} {\bibfield  {journal} {\bibinfo
   {journal} {Phys. Rev. B}\ }\textbf {\bibinfo {volume} {88}},\ \bibinfo
  {pages} {245403} (\bibinfo {year} {2013})}\BibitemShut {NoStop}%
\bibitem [{\citenamefont {Castellanos-Gomez}\ \emph {et~al.}(2014)\citenamefont
  {Castellanos-Gomez}, \citenamefont {Buscema}, \citenamefont {Molendaar},
  \citenamefont {Singh}, \citenamefont {Janssen}, \citenamefont {van~der
  Zant},\ and\ \citenamefont {Steele}}]{acg14}%
  \BibitemOpen
  \bibfield  {author} {\bibinfo {author} {\bibfnamefont {A.}~\bibnamefont
  {Castellanos-Gomez}}, \bibinfo {author} {\bibfnamefont {M.}~\bibnamefont
  {Buscema}}, \bibinfo {author} {\bibfnamefont {R.}~\bibnamefont {Molendaar}},
  \bibinfo {author} {\bibfnamefont {V.}~\bibnamefont {Singh}}, \bibinfo
  {author} {\bibfnamefont {L.}~\bibnamefont {Janssen}}, \bibinfo {author}
  {\bibfnamefont {H.~S.~J.}\ \bibnamefont {van~der Zant}}, \ and\ \bibinfo
  {author} {\bibfnamefont {G.~A.}\ \bibnamefont {Steele}},\ }\href@noop {}
  {\bibfield  {journal} {\bibinfo  {journal} {2D Materials}\ }\textbf {\bibinfo
  {volume} {1}},\ \bibinfo {pages} {011002} (\bibinfo {year}
  {2014})}\BibitemShut {NoStop}%
\bibitem [{\citenamefont {Plechinger}\ \emph {et~al.}(2014)\citenamefont
  {Plechinger}, \citenamefont {Mann}, \citenamefont {Preciado}, \citenamefont
  {Barroso}, \citenamefont {Nguyen}, \citenamefont {Eroms}, \citenamefont
  {Sch\"uller}, \citenamefont {Bartels},\ and\ \citenamefont
  {Korn}}]{Plechinger_SST14}%
  \BibitemOpen
  \bibfield  {author} {\bibinfo {author} {\bibfnamefont {G.}~\bibnamefont
  {Plechinger}}, \bibinfo {author} {\bibfnamefont {J.}~\bibnamefont {Mann}},
  \bibinfo {author} {\bibfnamefont {E.}~\bibnamefont {Preciado}}, \bibinfo
  {author} {\bibfnamefont {D.}~\bibnamefont {Barroso}}, \bibinfo {author}
  {\bibfnamefont {A.}~\bibnamefont {Nguyen}}, \bibinfo {author} {\bibfnamefont
  {J.}~\bibnamefont {Eroms}}, \bibinfo {author} {\bibfnamefont
  {C.}~\bibnamefont {Sch\"uller}}, \bibinfo {author} {\bibfnamefont
  {L.}~\bibnamefont {Bartels}}, \ and\ \bibinfo {author} {\bibfnamefont
  {T.}~\bibnamefont {Korn}},\ }\href@noop {} {\bibfield  {journal} {\bibinfo
  {journal} {Semiconductor Science and Technology}\ }\textbf {\bibinfo {volume}
  {29}},\ \bibinfo {pages} {064008} (\bibinfo {year} {2014})}\BibitemShut
  {NoStop}%
\bibitem [{\citenamefont {Peimyoo}\ \emph {et~al.}(2014)\citenamefont
  {Peimyoo}, \citenamefont {Yang}, \citenamefont {Shang}, \citenamefont {Shen},
  \citenamefont {Wang},\ and\ \citenamefont {Yu}}]{Peim14}%
  \BibitemOpen
  \bibfield  {author} {\bibinfo {author} {\bibfnamefont {N.}~\bibnamefont
  {Peimyoo}}, \bibinfo {author} {\bibfnamefont {W.}~\bibnamefont {Yang}},
  \bibinfo {author} {\bibfnamefont {J.}~\bibnamefont {Shang}}, \bibinfo
  {author} {\bibfnamefont {X.}~\bibnamefont {Shen}}, \bibinfo {author}
  {\bibfnamefont {Y.}~\bibnamefont {Wang}}, \ and\ \bibinfo {author}
  {\bibfnamefont {T.}~\bibnamefont {Yu}},\ }\href@noop {} {\bibfield  {journal}
  {\bibinfo  {journal} {ACS Nano}\ }\textbf {\bibinfo {volume} {8}},\ \bibinfo
  {pages} {11320} (\bibinfo {year} {2014})}\BibitemShut {NoStop}%
\bibitem [{\citenamefont {Bellus}\ \emph {et~al.}(2015)\citenamefont {Bellus},
  \citenamefont {Ceballos}, \citenamefont {Chiu},\ and\ \citenamefont
  {Zhao}}]{Bellus}%
  \BibitemOpen
  \bibfield  {author} {\bibinfo {author} {\bibfnamefont {M.~Z.}\ \bibnamefont
  {Bellus}}, \bibinfo {author} {\bibfnamefont {F.}~\bibnamefont {Ceballos}},
  \bibinfo {author} {\bibfnamefont {H.-Y.}\ \bibnamefont {Chiu}}, \ and\
  \bibinfo {author} {\bibfnamefont {H.}~\bibnamefont {Zhao}},\ }\href {\doibase
  10.1021/acsnano.5b02144} {\bibfield  {journal} {\bibinfo  {journal} {ACS
  Nano}\ }\textbf {\bibinfo {volume} {9}},\ \bibinfo {pages} {6459} (\bibinfo
  {year} {2015})}\BibitemShut {NoStop}%
\bibitem [{\citenamefont {Varshni}(1967)}]{Varshni_Physica67}%
  \BibitemOpen
  \bibfield  {author} {\bibinfo {author} {\bibfnamefont {Y.}~\bibnamefont
  {Varshni}},\ }\href@noop {} {\bibfield  {journal} {\bibinfo  {journal}
  {Physica}\ }\textbf {\bibinfo {volume} {34}},\ \bibinfo {pages} {149}
  (\bibinfo {year} {1967})}\BibitemShut {NoStop}%
\bibitem [{\citenamefont {Korn}\ \emph {et~al.}(2011)\citenamefont {Korn},
  \citenamefont {Heydrich}, \citenamefont {Hirmer}, \citenamefont
  {Schmutzler},\ and\ \citenamefont {Sch\"{u}ller}}]{Korn11}%
  \BibitemOpen
  \bibfield  {author} {\bibinfo {author} {\bibfnamefont {T.}~\bibnamefont
  {Korn}}, \bibinfo {author} {\bibfnamefont {S.}~\bibnamefont {Heydrich}},
  \bibinfo {author} {\bibfnamefont {M.}~\bibnamefont {Hirmer}}, \bibinfo
  {author} {\bibfnamefont {J.}~\bibnamefont {Schmutzler}}, \ and\ \bibinfo
  {author} {\bibfnamefont {C.}~\bibnamefont {Sch\"{u}ller}},\ }\href {\doibase
  10.1063/1.3636402} {\bibfield  {journal} {\bibinfo  {journal} {Applied
  Physics Letters}\ }\textbf {\bibinfo {volume} {99}},\ \bibinfo {eid} {102109}
  (\bibinfo {year} {2011})}\BibitemShut {NoStop}%
\bibitem [{sup()}]{supp}%
  \BibitemOpen
  \href@noop {} {}\bibinfo {note} {See Supplementary Information for further
  details}\BibitemShut {NoStop}%
\bibitem [{\citenamefont {Ye}\ \emph {et~al.}(2015)\citenamefont {Ye},
  \citenamefont {Wong}, \citenamefont {Lu}, \citenamefont {Zhu}, \citenamefont
  {Chen}, \citenamefont {Wang},\ and\ \citenamefont {Zhang}}]{Zhang15}%
  \BibitemOpen
  \bibfield  {author} {\bibinfo {author} {\bibfnamefont {Y.}~\bibnamefont
  {Ye}}, \bibinfo {author} {\bibfnamefont {Z.~J.}\ \bibnamefont {Wong}},
  \bibinfo {author} {\bibfnamefont {X.}~\bibnamefont {Lu}}, \bibinfo {author}
  {\bibfnamefont {H.}~\bibnamefont {Zhu}}, \bibinfo {author} {\bibfnamefont
  {X.}~\bibnamefont {Chen}}, \bibinfo {author} {\bibfnamefont {Y.}~\bibnamefont
  {Wang}}, \ and\ \bibinfo {author} {\bibfnamefont {X.}~\bibnamefont {Zhang}},\
  }\href@noop {} {\bibfield  {journal} {\bibinfo  {journal} {arXiv:1503.06141}\
  } (\bibinfo {year} {2015})}\BibitemShut {NoStop}%
\bibitem [{\citenamefont {Lee}\ \emph {et~al.}(2014)\citenamefont {Lee},
  \citenamefont {Lee}, \citenamefont {van~der Zande}, \citenamefont {Chen},
  \citenamefont {Li}, \citenamefont {Han}, \citenamefont {Cui}, \citenamefont
  {Arefe}, \citenamefont {Nuckolls}, \citenamefont {Heinz}, \citenamefont
  {Guo}, \citenamefont {Hone},\ and\ \citenamefont {Kim}}]{Kim14}%
  \BibitemOpen
  \bibfield  {author} {\bibinfo {author} {\bibfnamefont {C.-H.}\ \bibnamefont
  {Lee}}, \bibinfo {author} {\bibfnamefont {G.-H.}\ \bibnamefont {Lee}},
  \bibinfo {author} {\bibfnamefont {A.~M.}\ \bibnamefont {van~der Zande}},
  \bibinfo {author} {\bibfnamefont {W.}~\bibnamefont {Chen}}, \bibinfo {author}
  {\bibfnamefont {Y.}~\bibnamefont {Li}}, \bibinfo {author} {\bibfnamefont
  {M.}~\bibnamefont {Han}}, \bibinfo {author} {\bibfnamefont {X.}~\bibnamefont
  {Cui}}, \bibinfo {author} {\bibfnamefont {G.}~\bibnamefont {Arefe}}, \bibinfo
  {author} {\bibfnamefont {C.}~\bibnamefont {Nuckolls}}, \bibinfo {author}
  {\bibfnamefont {T.~F.}\ \bibnamefont {Heinz}}, \bibinfo {author}
  {\bibfnamefont {J.}~\bibnamefont {Guo}}, \bibinfo {author} {\bibfnamefont
  {J.}~\bibnamefont {Hone}}, \ and\ \bibinfo {author} {\bibfnamefont
  {P.}~\bibnamefont {Kim}},\ }\href@noop {} {\bibfield  {journal} {\bibinfo
  {journal} {Nat. Nanotechnol.}\ }\textbf {\bibinfo {volume} {9}},\ \bibinfo
  {pages} {676} (\bibinfo {year} {2014})}\BibitemShut {NoStop}%
\bibitem [{\citenamefont {Gourley}\ and\ \citenamefont
  {Wolfe}(1979)}]{Gourley}%
  \BibitemOpen
  \bibfield  {author} {\bibinfo {author} {\bibfnamefont {P.~L.}\ \bibnamefont
  {Gourley}}\ and\ \bibinfo {author} {\bibfnamefont {J.~P.}\ \bibnamefont
  {Wolfe}},\ }\href {\doibase 10.1103/PhysRevB.20.3319} {\bibfield  {journal}
  {\bibinfo  {journal} {Phys. Rev. B}\ }\textbf {\bibinfo {volume} {20}},\
  \bibinfo {pages} {3319} (\bibinfo {year} {1979})}\BibitemShut {NoStop}%
\bibitem [{\citenamefont {St\'{e}b\'{e}}\ \emph {et~al.}(1998)\citenamefont
  {St\'{e}b\'{e}}, \citenamefont {Feddi}, \citenamefont {Ainane},\ and\
  \citenamefont {Dujardin}}]{Stebe98}%
  \BibitemOpen
  \bibfield  {author} {\bibinfo {author} {\bibfnamefont {B.}~\bibnamefont
  {St\'{e}b\'{e}}}, \bibinfo {author} {\bibfnamefont {E.}~\bibnamefont
  {Feddi}}, \bibinfo {author} {\bibfnamefont {A.}~\bibnamefont {Ainane}}, \
  and\ \bibinfo {author} {\bibfnamefont {F.}~\bibnamefont {Dujardin}},\
  }\href@noop {} {\bibfield  {journal} {\bibinfo  {journal} {Phys. Rev. B}\
  }\textbf {\bibinfo {volume} {58}},\ \bibinfo {pages} {9926} (\bibinfo {year}
  {1998})}\BibitemShut {NoStop}%
\end{thebibliography}%
\onecolumngrid
\appendix
\subsection{Supplementary note 1: Reconstruction of the trion PL peak shape}
\begin{figure}[h]
\includegraphics[width=0.5 \linewidth]{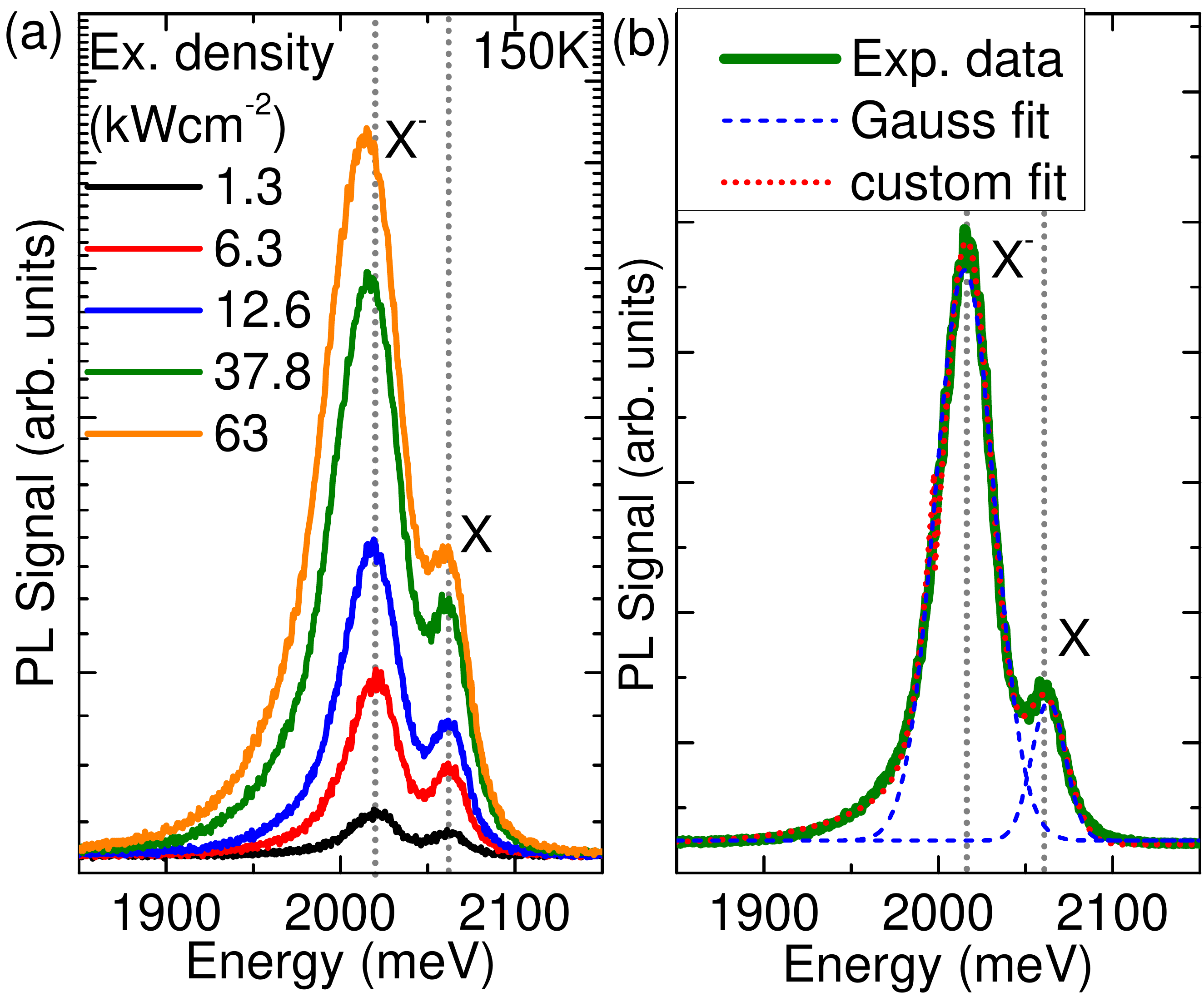}
\caption{(c) PL spectra of single-layer WS$_2$ at $T=150$\,K for different excitation densities.(b) PL spectrum of single-layer WS$_2$ at $T=150$\,K for an excitation density of 63~kWcm$^{-2}$ (green solid line). The data is fitted with a custom fit function including a double Gauss peak and an exponentially decaying low-energy shoulder for the X$^-$ peak (red dotted line). The blue dashed curves simulate two Gauss peaks.}
\label{fig:ElectronRecoil}
\end{figure}

In PL measurements at intermediate temperatures, we  observe  only the X and the X$^-$ peak without any contributions of biexcitons or low-energy defect-bound excitons, as the excitation-density-dependent measurements in Fig.~\ref{fig:ElectronRecoil}(a) show.
Close investigation of the PL peak shapes shows that the spectrum cannot be accurately described by the sum of two  Gaussian lineshapes, as indicated by the blue dashed lines in fig.~\ref{fig:ElectronRecoil}(b). Clearly, the trion peak has an asymmetrical shape with a  low-energy tail. In order to account for this feature, we have to consider electron recoil effects. Following Ref.~\cite{Ross13} and \cite{Stebe98}, the photon emission rate $R(\omega)$ is proportional to the convolution of the undisturbed, Gaussian-like emission rate $R(\omega_0)$, and an exponential decay at the low-energy side of the trion peak:
\begin{equation}R(\omega)=R(\omega_0)\exp\left[-\alpha(\hbar \omega_0 - \hbar \omega)\right]\Theta(\omega_0-\omega), \end{equation}
with $\omega_0$ being the trion frequency, $\alpha$ a parameter depending on temperature and effective mass, and $\Theta$ the Heaviside step function. In a first approximation to this assumption, we fit our experimental data with the following function:
\begin{equation}
I(E)=I_0 +A_1 \cdot G(E-E_{c1},w_1)+A_2 \cdot G(E-E_{c2},w_2))+A_3 \cdot \exp(\frac{E-E_{c3}}{t}) \cdot \Theta(E-E_{c3}),
\end{equation}
with $I(E)$ being the PL intensity as a function of the energy $E$, $A_{1,2}, w_{1,2}$ and $E_{c1,c2}$ the amplitude, width and center energy for the Gaussian peaks $G(E)$, $A_3$ the amplitude and $E_{c3}$ the onset energy for the low-energy exponential decay and $t$ the decay parameter.
The above equation fits well to our experimental data, as exemplarily indicated by the red dotted line in Fig.~\ref{fig:ElectronRecoil}(b), supporting our identification of the low-energy peak as the trion.
\subsection{Supplementary note 2: Trion-exciton peak separation in gate-dependent PL measurements}
\begin{figure}[h]
\includegraphics[width=\linewidth]{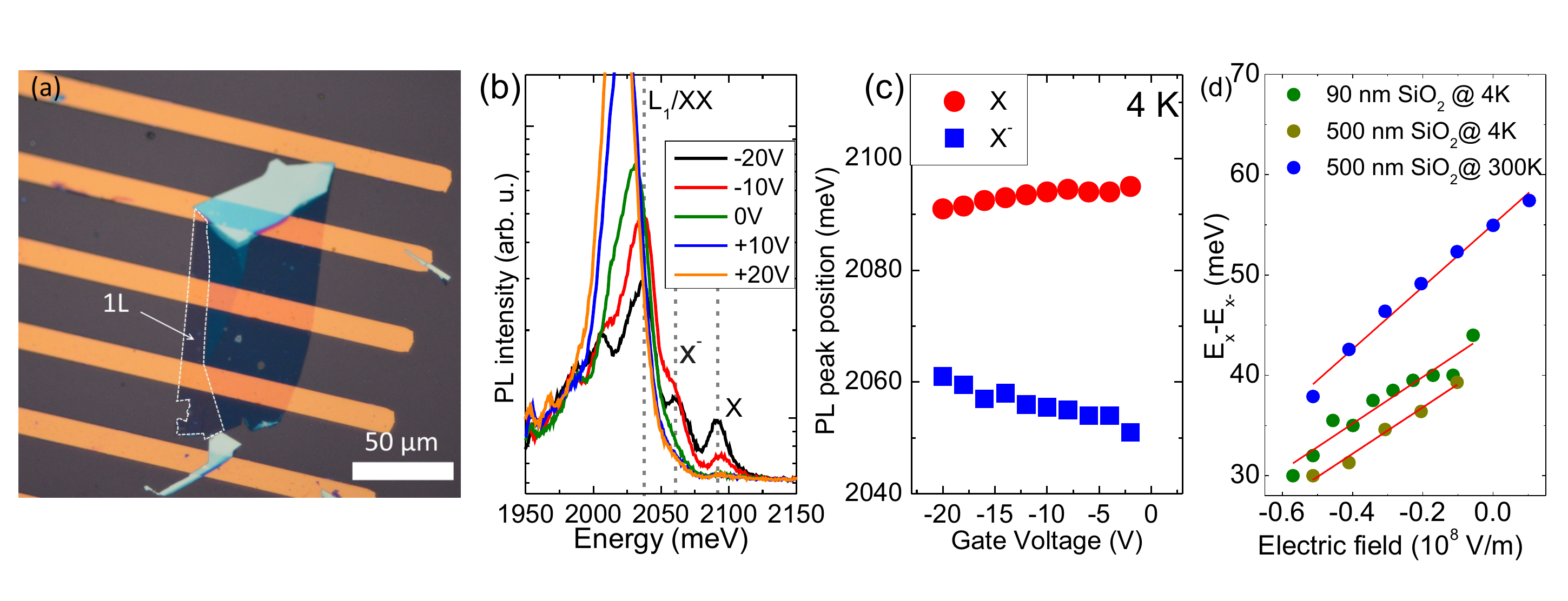}
\caption{(a) Optical micrograph of the WS$_2$ flake deposited on a Si chip with 90\,nm SiO$_2$ thermal oxide and prestructured Ti:Au stripes. (b) PL spectra at $T=4$\,K for different gate voltages. (c) PL peak position of X and X$^-$ as a function of gate voltage. (d) Energy separation of the X and X$^-$ PL peaks as function of the applied electric field for samples with different SiO$_2$ thickness at 4~K and room temperature. The red lines represent linear fits to the data.}
\label{fig:SuppGate}
\end{figure}
As discussed in the main text, we observe a clear dependence of the exciton-trion peak separation on the applied gate voltage in our samples due to a change of the Fermi energy~\cite{Mak13}. The results presented in the manuscript are confirmed by measurements on a second sample, in which  a p-doped Si chip with 90\,nm SiO$_2$ thermal oxide and prestructured Ti:Au stripes on top serves as the final substrate for our transfer process. Fig.~\ref{fig:SuppGate}(a) shows a microscope image of this sample. We use the Si as the backgate and observe a very similar behavior as discussed in the main text (see Fig.~\ref{fig:SuppGate}(b) and (c)). The X peak experiences a reduction of its intensity and a blueshift with increasing gate voltage, whereas the X$^-$ peak undergoes a redshift and an increase of its intensity. For positive gate voltages, the X$^-$ feature merges with the L$_1$ peak. By taking into account the oxide thickness and the dielectric constant of SiO$_2$, we get a measure for the electric field in our structures. Thus, we can directly compare samples with 90\,nm and 500\,nm SiO$_2$. Fig.~\ref{fig:SuppGate}(d) highlights the peak separation energy of exciton and trion, $E_X-E_{X^-}$, as a function of the applied electric field for different samples and temperatures. The data exhibits the expected linear correlation, as indicated by the linear fits. Both sample structures show a very similar behavior in PL measurements at 4~K, with a nearly identical linear slope. The vertical offset between the measurements on the samples indicates that the two contacted WS$_2$ flakes have different residual doping. The most likely cause for the vertical offset in the measurements on the sample with 500\,nm SiO$_2$ layer at different temperatures is the sample environment: the room-temperature measurement was performed with the sample in air, while low-temperature measurements were performed with the sample in vacuum. Surface adsorbates may accumulate on top of a flake under ambient conditions and shift the Fermi level.   The minimum peak separation between exciton and trion that we observe in gate-dependent measurements is 30\,meV, which is an upper boundary for the trion binding energy.

\end{document}